\documentstyle[12pt]{article}

\begin{document}

\title{Propagation of an optical pulse in a fiber link with random dispersion
management}
\author{Anders Berntson \\
Ericsson Telecom AB, S-12625 Stockholm, Sweden \and Boris A. Malomed
\\
Department of Interdisciplinary Studies, Faculty of Engineering, \\
Tel Aviv University, Tel Aviv 69978, Israel}
\maketitle

{\bf ABSTRACT}

A model of a long optical communication line consisting of alternating
segments with anomalous and normal dispersion, whose lengths are picked up
randomly from a certain interval, is considered. As the 
first stage of the
analysis, we calculate small changes of parameters of a quasi-Gaussian
pulse passing a double-segment cell by means of the variational
approximation (VA)
and approximate the evolution of the pulse passing many cells by smoothed
ODEs with random coefficients, which are solved numerically. Next, we
perform systematic direct simulations of the model. Results are presented as
dependences of the pulse's mean width, and standard deviation of the width
from its mean value, on the propagation distance. Averaging over $200$
different realizations of the random length set reveals slow long-scale
dynamics of the pulse, frequently in the form of long-period oscillations of
its width. It is thus found that the soliton is most stable in the
case of
the zero path-average dispersion (PAD), less stable in the case of anomalous
PAD, and least stable in the case of normal PAD. The soliton's stability
also strongly depends on its energy, the soliton with small energy
being much more robust than its large-energy counterpart.

\section{Introduction}

The use of dispersion management (DM) for improvement of pulse
transmission
in long optical-fiber links has attracted a great deal of attention,
see,
e.g., Refs. \cite{general}-\cite{interaction} and references therein.
Almost
all the works on this topic published thus far were dealing with
conditions
for stationary propagation of a pulse in a link consisting of
periodically
alternating fixed-length pieces of the fiber with anomalous and normal
dispersion. However, really exiting terrestrial optical communication
webs
are patchwork systems, which include links with different values of
the
dispersion and, what is especially important, very different lengths.
If the
distribution of the lengths may be assumed random, upgrading the
patchwork
systems by means of DM makes it necessary to consider transmission of
pulses
in long communication lines subject to {\em random} DM. Besides the
obvious
significance for the applications, this issue if also of considerable
interest by itself, in the context of soliton propagation in strongly
inhomogeneous media (see, e.g., the book 
\cite{Konotop}). The objective of
this work is to develop both semi-analytical variational and fully
numerical
approaches to the description of pulses' dynamics in a random-DM
system. The
variational approximation (VA), which was first applied to the DM
models in 
\cite{first}, has then become a commonly accepted tool for theoretical
consideration of DM solitons, see, e.g., Refs.
\cite{VA,Lakoba,Anders}.

Most promising results, viz., solitons that are most robust against
random
variation of the fiber-segment lengths, will be found in this work for
the
case of {\em zero} path-average dispersion (PAD) $\beta _{0}$. Note
that the
case of zero or slightly normal ($\beta _{0}>0$) PAD has recently
attracted
special attention in the context of regular (periodic) DM, after it
had been
demonstrated that stable propagation of pulses is possible in this
case too 
\cite{normal,Lakoba,Anders}.

There are two different natural models of random DM. In the {\it
two-lengths}
(2L) model, the lengths of the alternating anomalous- and
normal-dispersion
pieces are selected randomly from a certain interval {\em
independently}
from each other. In the other, {\it one-length} (1L) model, {\em
equal}
lengths of the anomalous- and normal-dispersion pieces inside each DM
cell
are selected randomly from the same interval. In the 2L model, $\beta
_{0}$
is the global average value, while in the 1L model $\beta _{0}$ is the
mean
dispersion in each cell. Final results are quite similar for both
models,
therefore in this work they will be presented for the 1L version.

Evolution of the soliton is a random setting is quantified by a
dependence
of its width on the propagation distance. In the random-DM links, the
soliton's width demonstrates rapid erratic oscillations. However,
averaging
over a large number of the random-DM realizations, and taking average
deviations from the mean values, we obtain smooth results which can be
understood. Roughly, the main conclusions are that robustness of a
randomly
dispersion-managed soliton crucially depends on the value of PAD and
on the
soliton's energy. First, the soliton's degradation (decay) is fast at
$\beta
_{0}>0$, slower at $\beta _{0}<0$ (recall this corresponds to the
normal and
anomalous dispersion, respectively), and {\em much slower} at $\beta
_{0}=0$%
. Second, degradation is much slower for solitons with small energy
than for
those with large energy. Both these findings can be understood if one
notices that, in the exactly solvable linear model (corresponding to
the
small-energy limit) with random DM, a chirped Gaussian pulse is
chaotically
vibrating with {\em no systematic degradation} (spreading out),
provided
that PAD is exactly zero (see, e.g., Ref. \ref{Lakoba}).

A preliminary version of the present work was published in
\cite{Dijon}.
Recently, another random-DM model was also considered \ by Abdullaev
and
Baizakov \cite{Tashkent} (see also \cite{Kepler}). However, in that
work
emphasis was made on a different model, in which the local values of
the
dispersion, rather than the fiber-segment lengths, were subject to a
random
distribution (the case of the randomly distributed lengths was only
briefly
touched within the framework of VA in Ref. \cite{Tashkent}). Unlike
the
results reported in the present work, sharp differences between robust
and
degrading soliton propagation regimes were not found in the model
considered
in Ref. \cite{Tashkent},

The rest of the paper is organized as follows. In section 2, the
random-DM
model is formulated. In section 3, we recapitulate basic results of VA
for
the periodic-DM model, and then develop VA for the random-DM one; the
approximation is based on the Gaussian {\it ansatz} for the pulse. In
section 4, results of systematic numerical simulations of the
VA-generated
equations with randomly varying coefficients are displayed. Section 5
presents results of {\em direct} numerical simulations of the
underlying
nonlinear Schr\"{o}dinger (NLS) equation with random DM, which are,
generally, found to be in accord with predictions of VA. Section 6
concludes
the paper.

\section{The random-dispersion-management model}

We start with a model which is a straightforward generalization of the
well-known DM schemes based on the NLS equation
\cite{general}-\cite{Lakoba}%
, 
\begin{equation}
iu_{z}-\frac{1}{2}\beta (z)u_{\tau \tau }+|u|^{2}u=0,  \label{NLS}
\end{equation}
which is written in the standard notation \cite{Agr}. The dispersion
coefficient is taken in the form $\beta (z)=\beta _{0}+\beta _{1}(z)$,
where 
$\beta _{0}$ is its average value (PAD), and $\beta _{1}(z)$ is a
variable
part with the zero average, which is taken in the following form
inside the $%
n$-th DM cell: 
\begin{equation}
\beta _{1}(z)\,=\,\left\{ 
\begin{array}{cl}
\beta {_{-}\,,} & z_{n}<z<z_{n}+L_{n}^{(-)}\,, \\ 
\beta _{+}{\,,} &
z_{n}+L_{n}^{(-)}<z<z_{n}+L_{n}^{(-)}+L_{n}^{(+)}\equiv
z_{n+1}\,.
\end{array}
\right.  \label{e4}
\end{equation}
$\,$Here, $\beta {_{-}}$ $+\beta _{0}<0$ and $\beta _{+}+\beta _{0}>0$
are,
respectively, the dispersion coefficient in the anomalous- and
normal-dispersion fibers and $L_{n}^{(\mp )}$ are lengths of the
corresponding pieces. In the case of periodic DM, the lengths
$L_{n}^{(\mp
)} $ are the same in all the cells, while in the case of random DM,
they
vary stochastically from a cell to a cell. The condition that the
average
value of $\beta _{1}(z)$ is zero implies that $\beta
_{-}\overline{L^{(-)}}%
+\beta _{+}\overline{L^{(+)}}=0$, the overbar standing for the
averaging. We
assume that random values of both $L^{(-)}$ and $L^{(+)}$ are
distributed
uniformly within a certain interval, so that their mean values are
equal.
Then, the above condition requires that $\left| \beta _{-}\right|
=\beta
_{+} $, which will be assumed to hold.

The model (\ref{NLS}) conserves the total energy $E$, which we define
as 
\begin{equation}
E\equiv \sqrt{2/\pi }\int_{-\infty }^{+\infty }|u(\tau )|^{2}d\tau .
\label{E}
\end{equation}
Following many earlier works \cite{general}-\cite{Greece}, the model
does
not include losses and gain, presuming that they are mutually
compensated at
a smaller scale. The analysis presented below can be extended to
include the
losses and gain, but this is postponed to another work, as it is
necessary
first of all to understand properties of the conservative model. Other
effects which are not included into the present work but may be
relevant in
certain cases, as it is known that they may help to stabilize the
transmission of pulses in the long periodic-DM link, are the
third-order
dispersion \cite{Greece} and filtering \cite{Filters}.

We are concerned with the case when DM is not too weak, hence the
shape of
the pulse is well-known to be close to a Gaussian \cite{general}-\cite
{Greece}. Accordingly, VA may be based on the Gaussian ansatz
\cite{VA}- 
\cite{Anders}. We will here follow a version of the Gaussian-based VA
developed (for periodic DM) in detail in \cite{Lakoba}. Using the
scaling
invariance of Eq. (\ref{NLS}), the following normalizations are
adopted in
Ref. \cite{Lakoba}, 
\begin{equation}
L^{(-)}+L^{(+)}\equiv 1,\ \left| \beta _{-}\right| L^{(-)}=\beta
_{+}L^{(+)}\equiv 1.  \label{Lakoba}
\end{equation}
In the present work, we apply these normalizations to the mean values
of the
random lengths. Since our model assumes
$\overline{L^{(-)}}=\overline{L^{(+)}%
}$, Eq. (\ref{Lakoba}) yields $\overline{L^{(\mp )}}=1/2$, and $\left|
\beta
_{\pm }\right| =2$. To comply with the former normalization, we choose
the
interval from which the random lengths $L^{(\mp )}$ are picked up as 
\begin{equation}
0.1<L<0.9\,.  \label{interval}
\end{equation}
The minimum length $0.1$ is introduced here because, in reality, the
length
can be neither very large (say, larger than $200$ km) nor very small
(shorter than $20$ km).

\section{The variational approximation}

\subsection{The Gaussian ansatz}

The strong-DM regime implies that, locally, the dispersion is much
stronger
than the nonlinearity, and that $\left| \beta _{0}\right| \ll \left|
\beta
_{\mp }\right| $. In the zero-order approximation, completely
neglecting the
nonlinearity, one has an exact Gaussian solution to the linearized
equation (%
\ref{NLS}) \cite{Lakoba}, 
\begin{equation}
u_{0}=\frac{\tau _{0}\sqrt{P_{0}}}{\sqrt{\tau _{0}^{2}+2i\Delta
(z)}}\,\exp %
\left[ -\frac{\tau ^{2}}{\tau _{0}^{2}+2i\Delta (z)}+i\phi \,\right]
\,.
\label{ourGauss}
\end{equation}
Here, $P_{0}$ and $\,\tau _{0}$ are, respectively, the peak power and
minimum width of the pulse, $-\,\Delta (z)$ $\equiv -\Delta
_{0}+\int_{z_{n}}^{z}\beta (z^{\prime })dz^{\prime }\;$is the {\it %
accumulated dispersion} defined inside the $n$-th DM cell, and
$\,\Delta
_{0} $ and $\,\phi $ are real constants.

The parameter $\tau _{0}^{-2}$ in the expression (\ref{ourGauss}) is
proportional, with regard to the normalizations (\ref{Lakoba}), to the
well-known {\em DM strength}, which (in the case of more general
normalizations) is defined as \cite{Anders} $S=\left( \left| \beta
_{-}\right| L^{(-)}+\beta _{+}L^{(+)}\right) /\tau _{{\rm FWHM}}^{2}$,
where 
$\tau _{{\rm FWHM}}$ is the full width at half maximum (FWHM) of the
pulse 
\cite{Agr} at the midpoint of the anomalous-dispersion segment, where
the
pulse is narrowest. Applying the notation adopted here, we replace
$L^{(\mp
)}$ in the definition of $S$ by the above mean values $1/2$, and also
insert 
$\left| \beta _{\pm }\right| =2$, obtaining 
\begin{equation}
S=1.443/\tau _{0}^{2}\,.  \label{Sour}
\end{equation}
The strength $S$ is the most important characteristic of the DM
schemes. It
determines their basic properties, which virtually do not depend on
other
parameters (such as, e.g., $L_{+}/L_{-}$), provided that $S$ is fixed.
In
particular, detailed numerical simulations reveal that stable DM
pulses do
not exist at $S>S_{{\rm \max }}\simeq 10$, the propagation at zero or
slightly normal PAD is possible is $S>S_{{\rm cr}}\approx 4$ \cite
{normal,Lakoba,Anders}, and strongest suppression of the interaction
between
the pulses is attained at $S\sim 1.6$ \cite{interaction}.

The exact solution (\ref{ourGauss}) for the linear model will be used
below
as an ansatz on which VA for the nonlinear model is based. In most
other
versions of VA \cite{VA}, the Gaussian {\it ansatz} is also used, but
in a
different form, 
\begin{equation}
u_{0}=a(z)\,\exp \left[ -\tau _{0}^{2}/W^{2}(z)+ib(z)\tau ^{2}+i\phi
\,%
\right] \,.  \label{otherGauss}
\end{equation}
The complex amplitude $\,a(z)$ and real width $\,W(z)$ and {\it chirp}
$%
\,b(z)$ introduced in this expression are related to parameters of the
ansatz (\ref{ourGauss}) as follows: 
\begin{equation}
a^{2}(z)=\tau _{0}^{2}P\left[ \tau _{0}^{2}+2i\Delta (z)\right]
^{-1}\,,\,\,\,W(z)=\tau _{0}^{-1}\sqrt{\tau _{0}^{4}+4\Delta ^{2}(z)}%
\,,\,\,\,b(z)=2\Delta \left[ \tau _{0}^{4}+4\Delta ^{2}(z)\right]
^{-1}\,.
\label{relation}
\end{equation}

Using the VA technique, one can derive equations for the
nonlinearity-induced evolution of the parameters $P$, $\tau _{0}$, and
$%
\Delta _{0}$, that were constant within the framework of the exact
Gaussian
solution in the absence of the nonlinearity \cite{Lakoba}. First, in
accord
with the energy conservation, we obtain $P_{0}\tau _{0}\equiv E={\rm
const}$
(this coincides with the conserved energy defined by Eq. (\ref{E}) ),
and
then 
\begin{equation}
\frac{d\tau _{0}}{dz}=\,\frac{\sqrt{2}E\tau _{0}\Delta
(z)}{W^{3}(z)}\,,\;%
\frac{d\Delta _{0}}{dz}=-\beta _{0}+\frac{E\,[4\Delta ^{2}(z)-\tau
_{0}^{4}]%
}{2\sqrt{2}\,W^{3}(z)}.  \label{evolution}
\end{equation}
Because the average dispersion is small in the DM regime, it is also
treated
(to derive the second equation in (\ref{evolution})) as a weak
perturbation.

The changes of the parameters $\tau _{0}$ and $\Delta _{0}$ per one DM
cell
can be calculated as 
\begin{equation}
\delta \tau _{0}=\oint \frac{d\tau _{0}}{dz}\,dz,\;\delta \Delta
_{0}=\oint 
\frac{d\Delta _{0}}{dz}\,dz,  \label{delta}
\end{equation}
where $\oint $ stands for the integration over a full cell, from
$z=z_{n}$
to $z=z_{n}+L_{n}^{(-)}\,+L_{n}^{(+)}$. In the spirit of the
perturbation
theory, the changes (\ref{delta}) are assumed small, hence,
calculating the
integrals in Eq. (\ref{delta}), it is sufficient to take into regard
only
the rapid variation of $\Delta (z)$, while $\tau _{0}$ and$\;\Delta
_{0}$
are treated as constants.

\subsection{Revisiting the case of the periodic dispersion management}

To develop VA for the random-DM system, it is first necessary to
recapitulate basic results for the usual periodic case. In that case,
to
obtain conditions providing for the stationary transmission of the
Gaussian
pulse in the long DM line, one equates to zero $\delta \tau _{0}$
and$%
\;\delta \Delta _{0}$, evaluated as per Eq. (\ref{delta}), with $\tau
_{0}$
and$\;\Delta _{0}$ kept constant inside the integrals. This yields
\cite
{Lakoba}

\begin{equation}
\Delta _{0}=-\frac{1}{2}\,,\;\frac{\beta
_{0}}{E}=\frac{\sqrt{2}}{4}\tau
_{0}^{3}\left[ \ln \left( \sqrt{1+\tau _{0}^{-4}}+\tau
_{0}^{-2}\right)
-2\left( \tau _{0}^{4}+1\right) ^{-1/2}\right] \,.  \label{stationary}
\end{equation}
In particular, Eqs. (\ref{stationary}) predict that the DM soliton
propagates steadily at anomalous average dispersion, $\beta _{0}<0$,
provided that $\,\tau _{0}^{2}>\left( \tau _{0}^{2}\right) _{{\rm cr}%
}\approx 0.301$, at $\beta _{0}=0$ if $\,\tau _{0}^{2}=\left( \tau
_{0}^{2}\right) _{{\rm cr}}$, and at {\em normal} average dispersion,
$\beta
_{0}>0$, if $\tau _{0}^{2}<\left( \tau _{0}^{2}\right) _{{\rm cr}}$.
$\left(
\tau _{0}^{2}\right) _{{\rm \min }}\approx 0.148$

In the case $\beta _{0}>0$, Eq. (\ref{stationary}) predicts that the
solution exists in a limited interval of the normal-PAD values, viz., 
\begin{equation}
0\leq \beta _{0}/E\leq \left( \beta _{0}/E\right) _{\max }\approx
0.0127.
\label{existence}
\end{equation}
Moreover, inside this interval Eq. (\ref{stationary}) yields {\em two}
different values of the minimum width $\tau _{0}$ for a given $\beta
_{0}/E$
(while in the anomalous-PAD region, $\tau _{0}$ is a uniquely defined
function of $\beta _{0}/E$) \cite{Lakoba}. On the basis of general
stability
criteria \cite{criteria}, one can immediately conclude that the
solution
(i.e., DM soliton) corresponding to the larger value of $\tau _{0}$ is
{\em %
stable}, while the one corresponding to the smaller $\tau _{0}$ is
{\em %
unstable}. The border between the stable and unstable solitons
corresponds
to $\beta _{0}/E=\left( \beta _{0}/E\right) _{\max }$, and it is at
$\tau
_{0}^{2}=\left( \tau _{0}^{2}\right) _{{\rm \min }}\approx 0.148$.
Translating $\tau _{0}^{2}$ into $S$ according to Eq. (\ref{Sour}) (in
particular, $\left( \tau _{0}^{2}\right) _{{\rm \min }}$ gives rise to
$%
S_{\max }\approx 9.75)$, we eventually conclude that VA based on Eqs.
(\ref{ourGauss}) and (\ref{evolution}) predicts the following:

\begin{itemize}
\item  stable DM solitons at anomalous path-average dispersion if
$S<S_{{\rm %
cr}}\approx 4.79;$

\item  stable DM solitons at {\em zero} path-average dispersion if
$S=S_{%
{\rm cr}}\approx 4.79;$

\item  stable DM solitons at {\em normal} path-average dispersion if
$%
4.79<S<S_{\max }\approx 9.75;$

\item  no stable DM soliton if $S>S_{\max }\approx 9.75$ .
\end{itemize}

Below, we will use, instead of $P_{0}$, the{\it \ }power normalized to
that
of the fundamental {\rm sech} soliton having the same width as a given
pulse 
\cite{Anders}, which is $P\equiv 4\cdot 1.12P_{0}$ (the factor $1.12$
is the
ratio of the FWHM for the {\rm sech}-shaped and Gaussian pulses). To
further
illustrate the properties of the solitons in the periodic-DM model, in
Fig.
1a we show the normalized power $P$ vs. the map strength $S$ for
different
values of $\beta _{0}$, as predicted by Eqs. (\ref{stationary}) and
(\ref
{Sour}) (the dependences are shown only in the region $S<9.75$, where
the
solitons are expected to be stable). For comparison, Fig. 1b shows the
same
dependence obtained from direct simulations of the full equation
(\ref{NLS}%
). The stars mark in Fig. 1 particular solutions whose response to
random
variations of the fiber segment lengths will be displayed below. The
curves
in Fig. 1b corresponding to normal PAD ($\beta _{0}>0$) terminate at
points
where the corresponding DM soliton becomes unstable.

The comparison of Figs. 1a and 1b shows that VA based on Eqs. (\ref
{ourGauss}) and (\ref{evolution}) yields quite acceptable results
(for
periodic DM) just in the range of small energies/powers, for which
this
approximation was devised. In particular, the VA-predicted $S_{{\rm
cr}%
}\approx 4.79$ is different from but nevertheless close to the
critical DM
strength $S_{{\rm cr}}\approx 4$ which the direct simulations give for
the
small-power case. With the increase of power, the numerically found
$S_{{\rm %
cr}}$ grows, as is evident in Fig. 1b. It is also noteworthy that the
value $%
S_{\max }\approx 9.75$, predicted by VA as the stability limit for the
DM
solitons, is indeed close to what is given by the direct simulations
for
small powers, see Fig. 1b.

At larger powers, there is a considerable discrepancy between VA and
the
direct numerical results. However, simulations displayed below show
that, in
the random-DM system, the soliton is strongly unstable at large
energies
anyway, so we are really interested only in the small-energy range,
for
which the above VA is adequate.

\section{The randomly dispersion-managed system}

In the general case, when the pulse transmission is not steady
(including
the random-DM case), its evolution from a cell to a cell can be
described in
terms of a {\it map}, $\tau _{0}\rightarrow \tau _{0}+\delta \tau
_{0}$, $%
\Delta _{0}\rightarrow \Delta _{0}+\delta \Delta _{0}$. Because the
changes
are small, many iterations of the map, corresponding to the
propagation
distance comprising many DM cells, may be approximated by smoothed
differential equations, $d\tau _{0}/dz=\delta \tau _{0}/\left(
L_{n}^{(-)}\,+L_{n}^{(+)}\right) $and $d\Delta _{0}/dz=\delta \Delta
_{0}/\left( L_{n}^{(-)}\,+L_{n}^{(+)}\right) $. A straightforward
calculation, using Eqs. (\ref{delta}) and (\ref{evolution}) and taking
into
regard the normalization $\left| \beta _{-}\right| =\left| \beta
_{+}\right|
=2$ adopted above, leads to a final form of the smoothed equations: 
\begin{eqnarray}
\frac{d\tau _{0}}{dz} &=&\frac{\sqrt{2}E\tau _{0}^{4}}{8\left[
L^{(-)}\,+L^{(+)}\right] }\,\{\frac{1}{\sqrt{\tau _{0}^{4}+4\Delta
_{0}^{2}}}%
+\frac{1}{\sqrt{\tau _{0}^{4}+4\left[ \Delta
_{0}+2L^{(-)}\,-2L^{(+)}\right]
^{2}}}  \nonumber \\
&&-\frac{2}{\sqrt{\tau _{0}^{4}+4\left[ \Delta _{0}+2L^{(-)}\,\right]
^{2}}}%
\}\,,  \label{tau0} \\
\frac{d\Delta _{0}}{dz} &=&-\beta _{0}+\frac{\sqrt{2}E\tau
_{0}^{3}}{8\left[
L^{(-)}\,+L^{(+)}\right] }\,\{\frac{2\Delta _{0}}{\sqrt{\tau
_{0}^{4}+4\Delta _{0}^{2}}}+\frac{2\left[ \Delta
_{0}+2L^{(-)}\,-2L^{(+)}%
\right] }{\sqrt{\tau _{0}^{4}+4\left[ \Delta
_{0}+2L^{(-)}\,-2L^{(+)}\right]
^{2}}}  \nonumber \\
&&-\frac{4\left[ \Delta _{0}+2L^{(-)}\right] }{\sqrt{\tau
_{0}^{4}+4\left[
\Delta _{0}+2L^{(-)}\right] ^{2}}}-\frac{1}{2}\ln \left( 2\Delta
_{0}+\sqrt{%
\tau _{0}^{4}+4\Delta _{0}^{2}}\right)  \nonumber \\
&&-\frac{1}{2}\ln \left( 2\left[ \Delta
_{0}+2L^{(-)}\,-2L^{(+)}\right] +%
\sqrt{\tau _{0}^{4}+4\left[ \Delta _{0}+2L^{(-)}\,-2L^{(+)}\right]
^{2}}%
\right)  \nonumber \\
&&+\ln \left( 2\left[ \Delta _{0}+2L^{(-)}\right] +\sqrt{\tau
_{0}^{4}+4%
\left[ \Delta _{0}+2L^{(-)}\right] ^{2}}\right) \}\,.  \label{Delta0}
\end{eqnarray}
To check these equations, one can get back to the case of periodic DM,
with $%
L^{(-)}\,=L^{(+)}=1/2$, as per Eqs. (\ref{Lakoba}). In this case, a
{\it %
fixed point} of Eqs. (\ref{tau0}) and (\ref{Delta0}) ($d\tau
_{0}/dz=d\Delta
_{0}/dz=0$) has exactly the same values of $\Delta _{0}$ and $\tau
_{0}$ as
given by Eqs. (\ref{stationary}). In the next section, we will display
results of numerical integration of Eqs. (\ref{tau0}) and
(\ref{Delta0}) for
the 1L model of random DM defined above, in which the values $%
L^{(+)}\,=L^{(-)}$ are picked up randomly from the interval
(\ref{interval}).

\section{Numerical simulations of the variational equations}

Equations (\ref{tau0}) and (\ref{Delta0}) with random coefficients
were
numerically integrated, with initial conditions corresponding to the
chirpless pulse, whose parameters were taken as per the fixed point
(\ref
{stationary}) of the allied periodic-DM model. The most essential
single
characteristic of the pulse propagation at given values of $\beta
_{0}$ and $%
E$ is the rms cell-average pulse's width $\overline{W}$, which we
define as 
\begin{equation}
\overline{W}\equiv L^{-1}\oint W(z)dz  \label{W}
\end{equation}
where the relation (\ref{relation}) and normalizations (\ref{Lakoba})
have
been used. It is noteworthy that, in the case of periodic DM, the
steady
propagation regime corresponding to the fixed point (\ref{stationary})
with $%
\Delta _{0}=-1/2$ gives\ rise to the minimum rms width at $\tau
_{0}^{2}=1/%
\sqrt{3}$ (i.e., in the anomalous-PAD region, as $1/\sqrt{3}>\left(
\tau
_{0}^{2}\right) _{{\rm cr}}\approx 0.30$).

Simulations of Eqs. (\ref{tau0}) and (\ref{Delta0}) reveal that there
are
two different dynamical regimes. In the case when the soliton's energy
is
sufficiently low, i.e., one is indeed close to the quasi-linear
approximation for which the above derivation is relevant, and PAD is
anomalous or zero, i.e., $\beta _{0}\leq 0$ (especially, if $\beta
_{0}=0$),
the pulse performs random vibrations but remains, as a matter of fact,
fairly stable over long propagation distances. In the case when the
energy
is higher, as well as when PAD is normal, $\beta _{0}>0$, the pulse
demonstrates fast degradation and spreading out.

To present the results in a more systematic form, in Fig. 2 we display
the
results for a typical case of the effectively stable propagation with
small
normalized power of the soliton, $P=0.16$ (in this case, $E=0.5$), and
anomalous PAD, $\beta _{0}=+0.2$ (which is $10\%$ of the local
dispersion).
The simulations of Eqs. (\ref{tau0}) and (\ref{Delta0}) were performed
$200$
times, with the same initial conditions but different realizations of
the
random length set. Using numerical data for the $200$ runs, we
computed the
evolution of $\left\langle \tau _{0}\right\rangle $ and the averaged
dependence $\left\langle \overline{W}(z)\right\rangle $ ($\left\langle
...\right\rangle $ stands for the averaging over $200$ runs), along
with the
corresponding standard deviations. Both dependences are displayed in
Fig. 2.

It is clearly seen in Fig. 2 and in a number of similar plots not
displayed
here that, on top of the random vibrations, which are directly induced
by
random DM but are eliminated by averaging over $200$ realizations, the
soliton demonstrates slow (long-scale) dynamics. Systematic
degradation of
the oscillations and of the soliton itself takes place too, the
degradation
being slower for lower powers. For the particular case shown in Fig 2,
the
pulses remain certainly usable over the propagation distance $\simeq
100$ DM
cells. Further simulations of Eqs. (\ref{tau0}) and (\ref{Delta0}) for
propagation distances essentially larger than $1000$ DM\ cells (not
shown
here; see an example in Ref. \cite{Dijon}) show that, in fact, the
sluggish
spreading out of the soliton suddenly ends up with its blowup
(complete
decay into radiation). It is interesting to note that VA has predicted
the
same scenario for the evolution of the soliton in the early work
\cite{first}
for a model with regular sinusoidal modulation of the local
dispersion: a
long span of chaotic but nevertheless quasi-stable vibrations is
suddenly
changed by rapid irreversible decay.

Fig. 3 shows a drastic difference in the soliton's evolution which
takes
place if the power is increased to $P=0.44$, (the corresponding energy
is $%
E=2.5$), without changing parameters of the random-DM fiber link. In
this
case, which is typical for high powers, rapid decay of the soliton
without
long-scale vibrations is observed. Note that, since $\left\langle \tau
_{0}\right\rangle $ grows quite slowly, and the spectral width of the
pulse
is $\sim \tau _{0}^{-1}$, most of the pulse's broadening in the
temporal
domain observed in Fig 3 is due to chirping of the pulse, rather than
directly to a change in its spectral width.

Taking $\beta _{0}=0$, instead of anomalous PAD, {\em radically
improves}
the situation, as is seen in Fig. 4. For the high-power case with
$P=0.44$ ($%
E=3.6$), which gave rise to rapid degradation of the soliton in the
presence
of the anomalous PAD, the pulse now survives over much longer
distances, see
Fig 4a. For lower values of the power, we observer still more robust
propagation with zero PAD. Fig 4b shows the case $P=0.12$ ($E=0.1$),
where
the pulse shows slow dynamics induced by the random length variation
in a
strongly dispersion-managed link ($S\approx 4.8$) with very little
degradation over 1000 DM cells.

The PAD's value $\beta _{0}=0$ turns out to be a point of a {\em sharp
optimum}: taking any tangible small normal value of PAD, $\beta
_{0}>0$, we
always observed rapid decay (without long-scale oscillations) of the
soliton
at virtually all the values of the energy, see for instance Fig 5,
which
displays the case with $P=0.27$ ($E=2.5$) and $\beta _{0}=0.02$ (1\%
of the
local dispersion). In this case the soliton character of the pulse is
lost
quickly. The broadening of the pulse in the spectral domain is
enhanced by
its spectral broadening, manifested in the decrease of $\tau _{0}$.
The
broadening rate, however, is slower than in Fig 3a due to the lower
magnitude of PAD.

\section{Direct simulations}

It is necessary to compare the predictions of VA with direct
simulations of
Eq. \ref{NLS}. The pattern of the simulations was the same as in the
previous section, i.e., simulations were performed for $200$ different
realizations of the random length set, in order to evaluate averaged
evolution of the pulse's parameters. First, the shape of the DM
soliton in
the allied periodic-DM link was numerically determined, to be used as
the
initial configuration. The randomness was the same as above, i.e., the
lengths were uniformly distributed between 0.2 and 1.8 of the average
length. The general trends predicted by the VA are confirmed by the
direct
simulations: lower energy and anomalous or, especially, zero PAD\
enhance
the stability, see details below.

In direct simulations, the width must be defined with special care.
The
usual FWHM definition may be misleading in the case of random DM, as
the
soliton can sometimes split (see below); besides that, this definition
ignores the presence of a radiation component. Therefore, we adopted
an {\em %
integral} definition, with which the width is a size of the temporal
region
on both sides of the soliton's center that contains $76\%$ of the net
energy. For a pulse with the Gaussian shape the width defined this way
coincides with FWHM. We will assume that optimum dispersion
compensation can
be applied at the receiver, i.e., any linear chirp is removed from the
pulse
by a dispersion compensating element installed before the receiver.
Note
that in the framework of the above VA, $\tau _{0}$ represented the
pulse's
width at the chirp-free points.

Starting with anomalous PAD, in Fig 6 we show results for the same
case for
which results obtained by means of VA were displayed in Fig 2, i.e.,
the
magnitude of the PAD is 10\% of the local dispersion and the power is
low, $%
P=0.18$. This time, we show (by bold curves) not only the averaged
evolution, but also (by thin curves) the set of the particular
evolutions
corresponding to different realizations of the random length set.
Comparing
Figs. 6 and 2, we conclude that the averaged results are qualitatively
similar. In particular, internal vibrations are present in the first
100
cells in both cases, and the systematic temporal-domain broadening of
the
pulse takes place in both cases, although the broadening rate is
overestimated by VA (hence, the full numerical results, predicting
{\em %
slower} degradation of the soliton, seem considerably {\em better} for
the
applications than the less accurate results generated by VA). The
latter
discrepancy is, most plausibly, accounted for by radiative losses that
are
ignored by the VA.

Since we allow for dispersion compensation at the receiving end, this
can
change the link's PAD. The effective PAD, corrected with regard to the
optimal receiving-end dispersion compensator, may therefore be
regarded as a
function of the propagation distance. The inset to Fig. 6 shows that
the
thus redefined effective PAD stabilizes at the value $0.2$, which
turns out
to be the same as for the unperturbed DM soliton. This means that no
{\it %
post-transmission} dispersion compensation is needed in \ order to
receive a
chirp-free pulse; the dispersion of the link is balanced by its
nonlinearity. For higher powers, however, this is no longer the case,
see an
example for $P=0.47$ in Fig. 7. The pulse broadens and decays fast and
the
optimum end dispersion drops rather than stabilizing, i.e., the pulse
acquires a chirp during the propagation and extra post-transmission
dispersion compensation will be necessary. Compared to the variational
results, displayed for the same case in Fig 3, we conclude that the
direct
simulations of Eq. (\ref{NLS}) reveal a rapid decay of the pulse
shape.
Moreover, detailed e4xamination of the pulse shapes generated by the
direct
simulations shows they often develop a multi-peak structure, so that
it can
sometimes be hard to identify the pulse proper. This is why Fig. 7
shows the
evolution for the first 50 DM cells only.

It is necessary to mention that essentially longer direct simulations
of Eq.
(\ref{NLS}) (not shown here) demonstrate that the gradual decay of the
soliton is, at a final stage, suddenly interrupted by {\em splitting
}of the
residual pulse into two smaller ones. Recall that, although VA by
definition
cannot predict the eventual sudden splitting of the pulse, it does
predict,
as it was mentioned above, something similar, viz., sudden fast decay
of the
soliton after a long stage of chaotic vibrations. It is noteworthy
that
gradual evolution of the soliton in the model with sinusoidal periodic
modulation of the dispersion also ends up with sudden splitting, which
approximately corresponds to the sudden decay predicted by VA in the
same
model \cite{Scripta,Abdullaev}.

To further illustrate the internal dynamics of the pulse, we picked a
case
where PAD is anomalous and equal 1\% of the local dispersion, and the
power
is $P=0.1$. Fig. 8 shows the average trajectory on the dynamical
plane,
where the coordinates are the width and chirp of the soliton at the
midpoint
of the anomalous-fiber section. The chirp is in this case defined as
the
second derivative of the phase of the pulse taken at its center. The
trajectory demonstrates a quasi-circular motion around a center which
is
drifting to the right, towards a broader pulse. The trajectory clearly
demonstrates the nonlinear character of the system, cf. the inset to
Fig. 8
which shows the trajectory if the nonlinearity are dropped. Note that
a
stationary periodic-DM soliton would be represented by a single point
\ in
this plane, and a perturbed (nonstationary) soliton in the periodic-DM
system would trace a circle around a fixed center shifted to a broader
pulse, as compared to the stationary DM soliton \cite{Dijon:slow}. The
random length variation is a permanently acting perturbation \ which
generates the persistent drift of the center in Fig. 8.

Numerical simulations at zero PAD also agree well with the results
generated
by VA: zero PAD provides the slowest pulse degradation in the random
DM
link. Fig. 9 shows the evolution of the pulse width when $\beta
_{0}=0$ and
the power is $P=0.15.$ Almost {\em no broadening} ($<5\%$) can be
observed,
on the average, after having passed $200$ DM cells. The broadening
does
increase with the power, but remains, as predicted by VA, much less
than in
the case of nonzero PAD. For instance, taking $P=0.45$ results only in
a
doubling of the width on the propagation distance equal 200 DM cells.
Comparison to the variational results for the same case displayed in
Fig. 4
again shows that full numerical results are qualitatively similar but
somewhat {\em better}, predicting slower degradation of the pulse. One
of
the reasons for this may be that the radiation remains trapped within
the
pulse.

Lastly, for the case of normal PAD the direct simulations confirm the
prediction of VA according to which the solitons are least stable in
this
case. This is partly explained by the fact that there is, effectively,
a
minimum (threshold) power necessary for the existence of stable
solitons at
normal PAD even with periodic DM, see Fig 1b, and the larger power
always
stimulates the degradation of the pulse. Detailed examination of the
direct
numerical results for the normal-PAD\ case also shows that the pulse's
spreading out is boosted by its spectral broadening during the initial
stage
of propagation. As a result, even at very weak normal PAD, $\beta
_{0}=0.02$
(1 \% of the magnitude of the local dispersion), the pulse stays
intact for
no more than 10 DM cells.

\section*{Conclusion}

In this work, we have put forward a model of a long optical
communication
line subject to random dispersion management (DM). The line consists
of
alternating fiber pieces with anomalous and normal dispersion, whose
lengths
are picked up randomly from a certain interval, while the absolute
values of
the dispersion coefficients in both pieces are always equal. By means
of the
variational approximation, we calculated small changes of parameters
of the
propagating quasi-Gaussian pulse per one DM cell, and then
approximated the
evolution of the pulse passing many cells by smoothed ordinary
differential
equations with randomly varying coefficients. The equations were
solved
numerically, and results were presented as the dependences of the
pulse's
mean width, and average deviation of the width from the mean value, on
the
propagation distance. Averaging the results over $200$ different
realizations of the random length set removes rapid oscillations of
the
width and reveals slow long-scale dynamics of the pulse, frequently in
the
form of long-period oscillations. The second part of the work is based
on
direct numerical simulation of the same model. Comparing the results,
we
have concluded that essential features of the soliton dynamics are the
same
in the variational approximation and in direct numerical simulations:
the
propagating soliton is most stable in the case of zero path-average
dispersion (PAD), less stable in the case of anomalous PAD, and least
stable
in the case of normal PAD. The soliton's stability also strongly
depends on
its energy, so that the soliton with small energy is much more robust
than
its large-energy counterpart.

\section{Acknowledgments}

We appreciate valuable discussions with M. Tur (Tel Aviv University),
M.
B\"{o}hm and F. Mitschke (University of Rostock), and F. Lederer
(University
of Jena).

\newpage

\section*{FIGURE CAPTIONS}

Fig. 1. The normalized power vs. the map strength for stationary
solitons in
the periodic DM system: (a) the analytical result, Eq.
(\ref{stationary}),
produced by the variational approximation; (b) direct numerical
solution of
the NLS equation (\ref{NLS}). Each line corresponds to a constant
value of
PAD: from left to right, $\beta _{0}=-0.2,-0.02,0$ and $0.02$. The
stars
mark particular cases for which response to random length variations
are
further displayed in Figs. 2 through 9.

Fig. 2. The cell-average pulse width (top) and minimum-width parameter
$\tau
_{0}$ (bottom) vs. the propagation distance, generated by numerical
integration of the variational equations (\ref{tau0}) and
(\ref{Delta0})
for the power $P=0.16$ and $\beta _{0}=-0.2$ (anomalous PAD). The mean
values (solid curve) and standard deviations from them (dashed curves)
are
produced by averaging over $200$ different realizations of the random
length
set.

Fig. 3. The same as in Fig. 2 but for higher power, $P=0.44$.

Fig. 4. Evolution of the cell-average pulse width for zero PAD
propagating
over 1000 DM cells, in the cases of high power $P=0.47$ (top) and low
power $%
P=0.1$ (bottom). The mean values (solid curve) and standard deviations
from
them (dashed curves) are produced by averaging over $200$ different
realizations of the random length set.

Fig. 5. The same as Fig. 2 but with normal PAD, $\beta _{0}=0.02$, and
$%
P=0.27.$

Fig. 6. The pulse width vs. the propagation distance for anomalous
PAD, $%
\beta _{0}=-0.2$, and low power, $P=0.18$, generated by direct
simulations
of Eq. (\ref{NLS}). Shown are both the mean values (solid curve) and
standard deviations from them (dashed curves) as produced by averaging
over $%
200$ different realizations of the random length set, and grey curves
corresponding to particular realizations of the random set. The inset
shows
the effective PAD corrected with regard to the optimal dispersion
compensation at the receiving edge (average of 200 simulations).

Fig. 7. The same as in Fig 6. but for high power, $P=0.47$.

Fig. 8. Results of direct simulations of the NLS equation (\ref{NLS})
with
anomalous PAD, $\beta _{0}=-0.02$. The plot shows the pulse dynamics
in the
dynamical plane whose coordinates are the chirp at the center of the
pulse
and its width, evaluated at the midpoint of the each anomalous-fiber
section. The inset shows the same but in the absence of the nonlinear
term
in Eq. (1).

Fig. 9. The same as in Fig. 6 for the case of zero PAD and $P=0.15$.

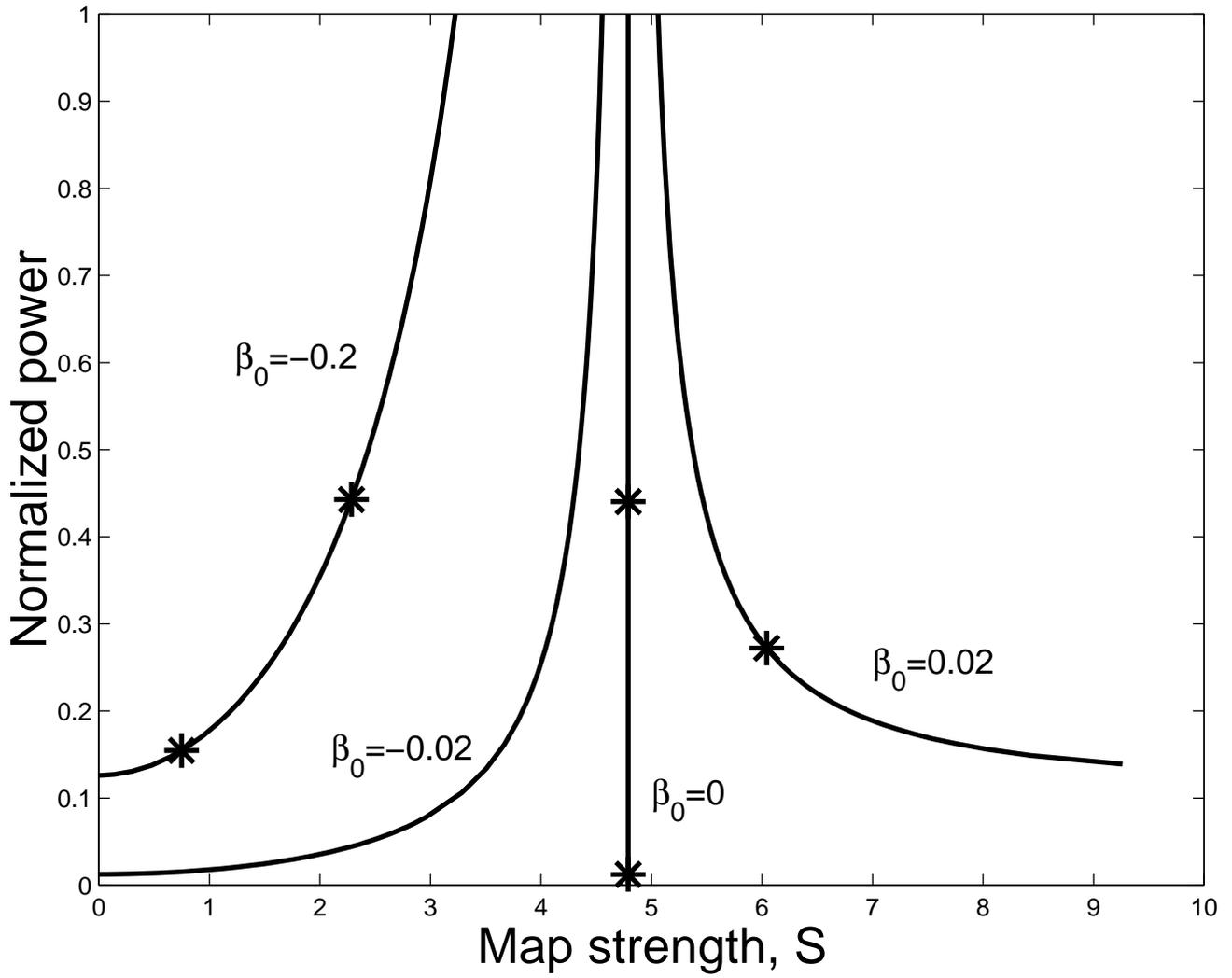
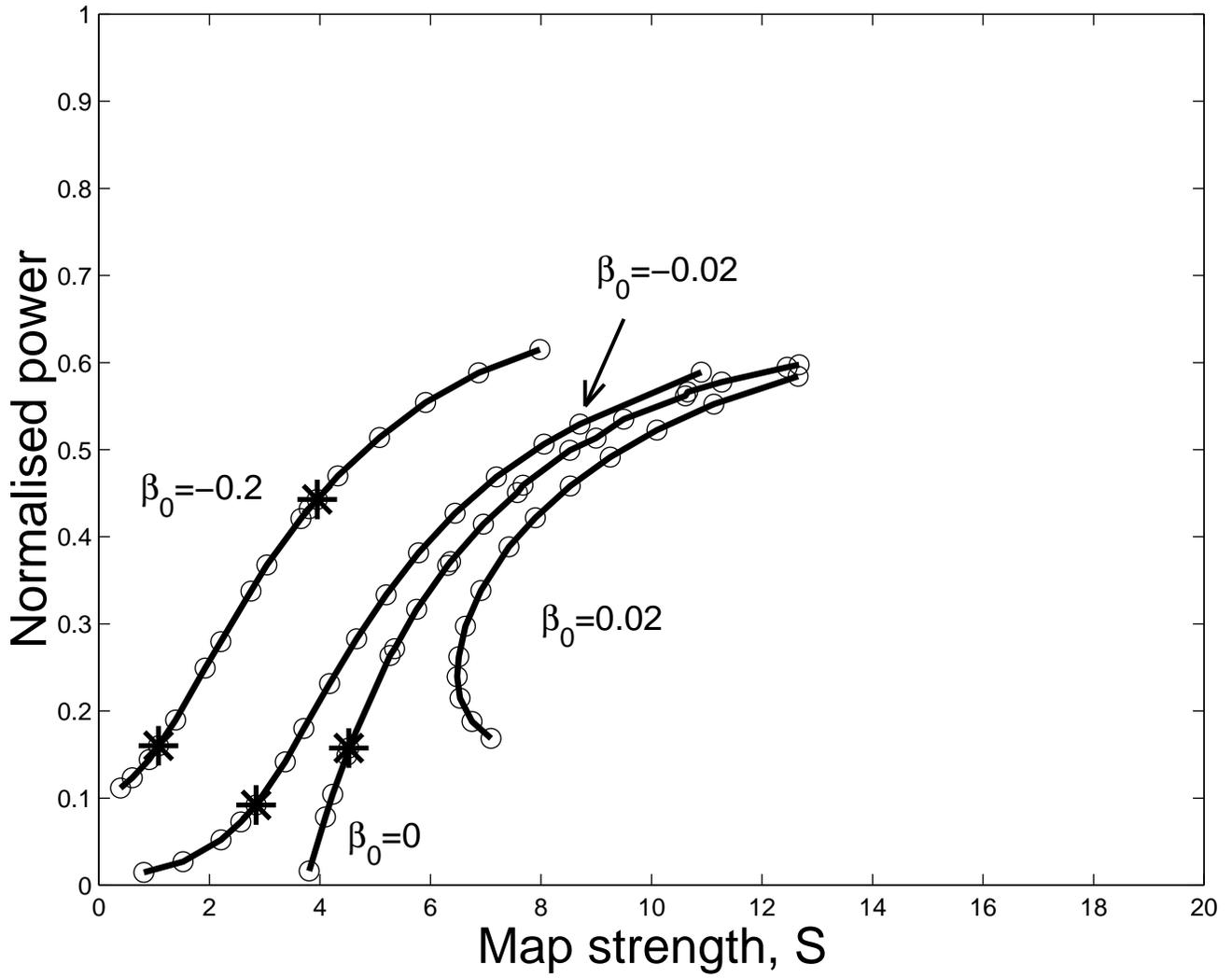
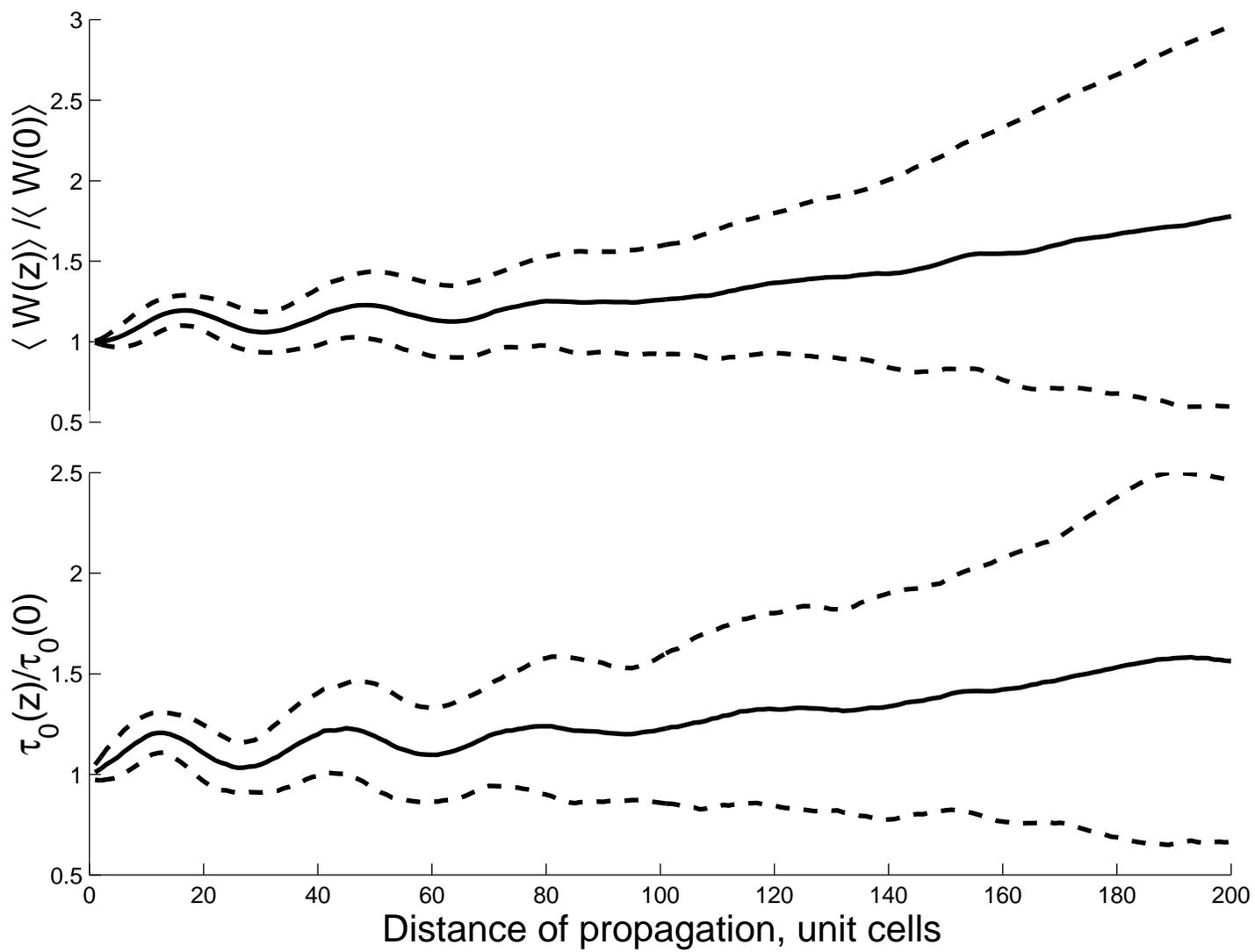
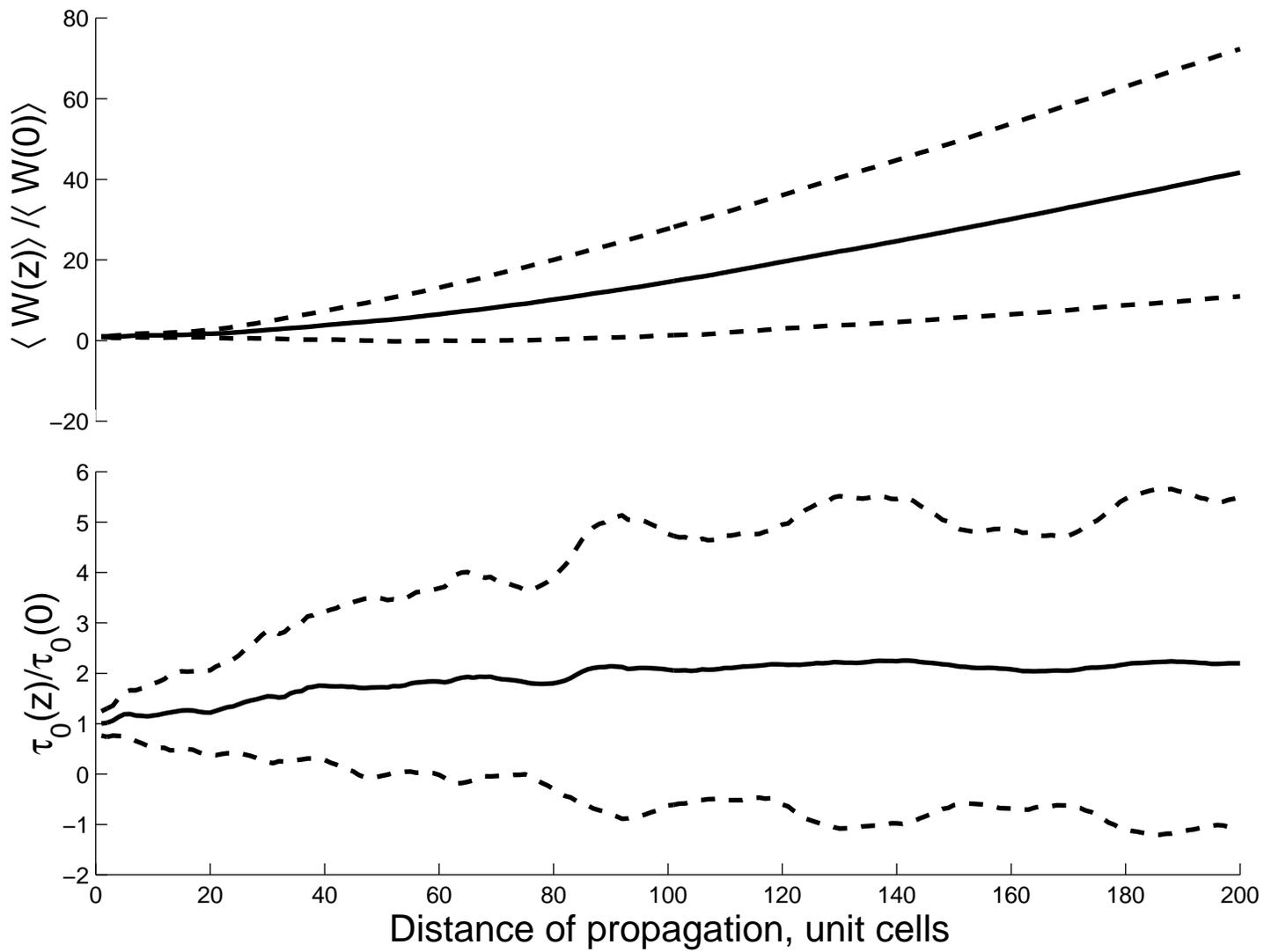
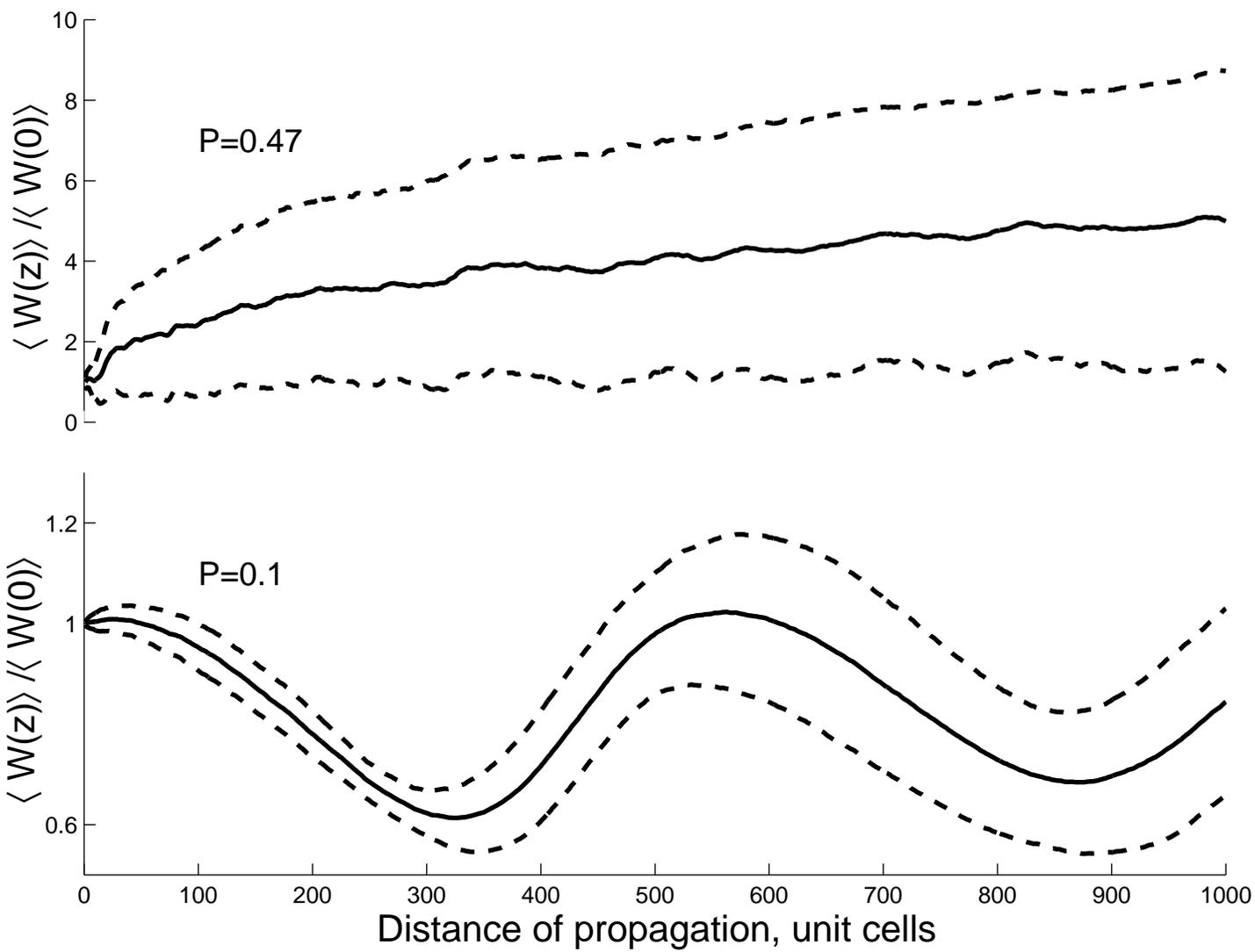
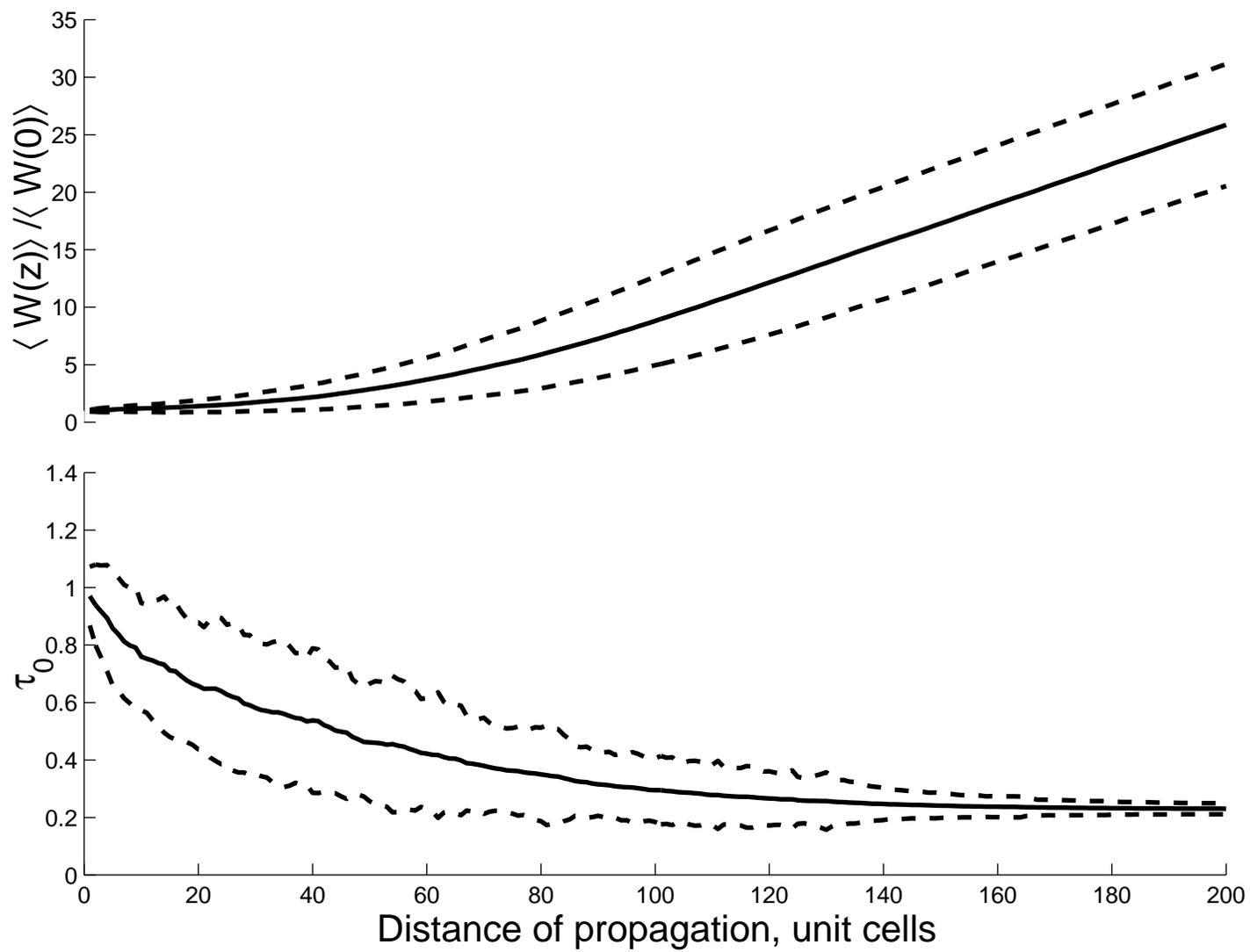
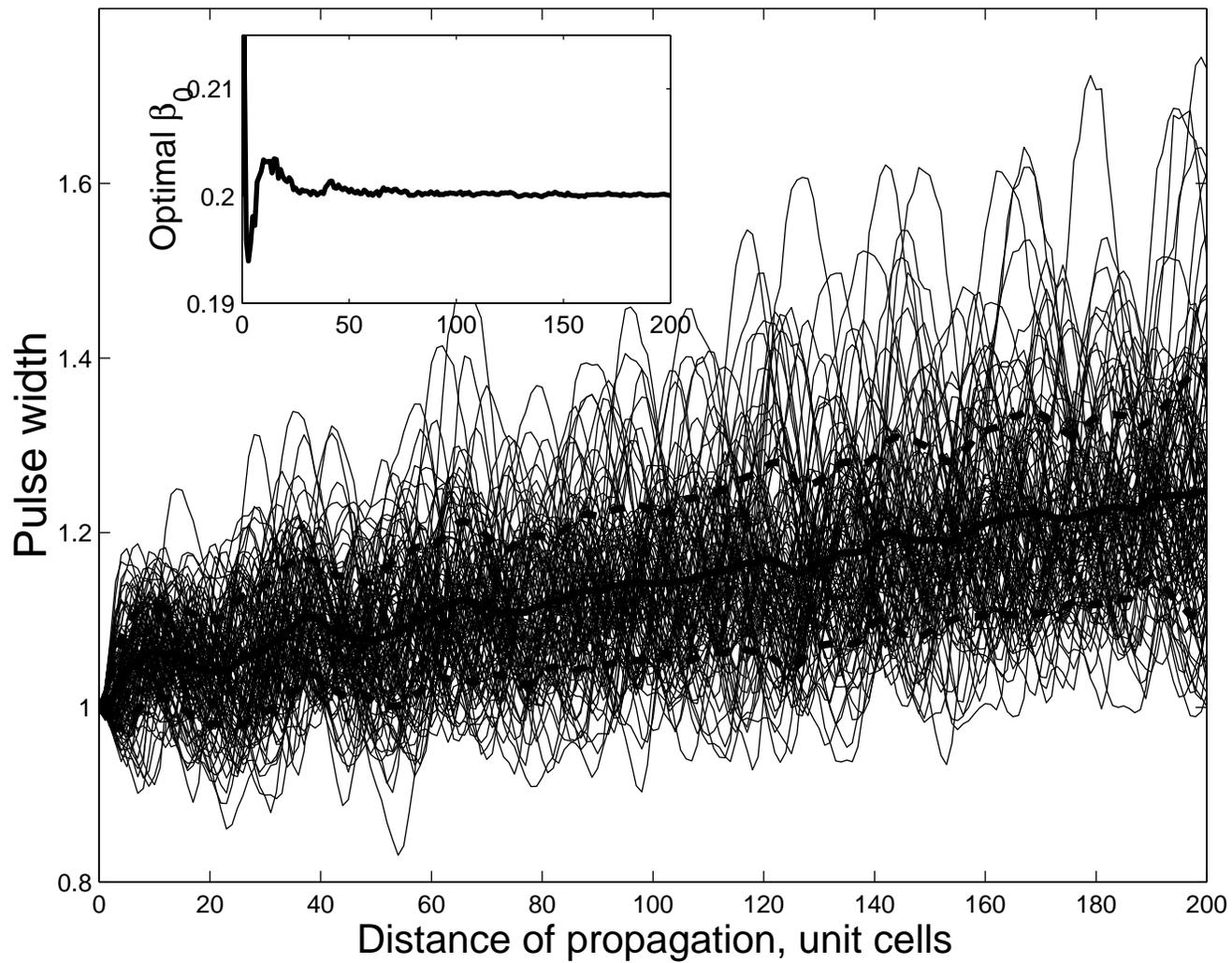
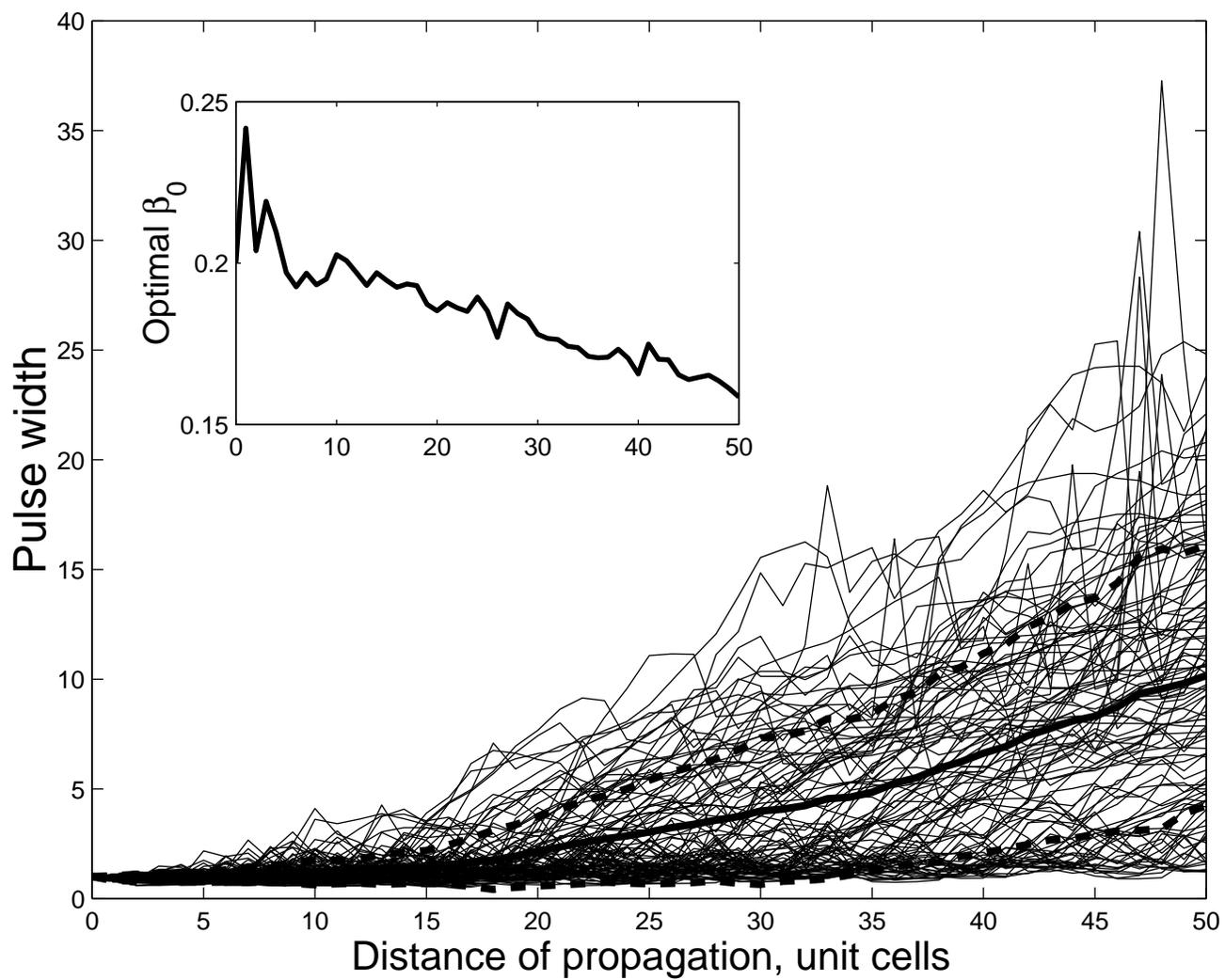
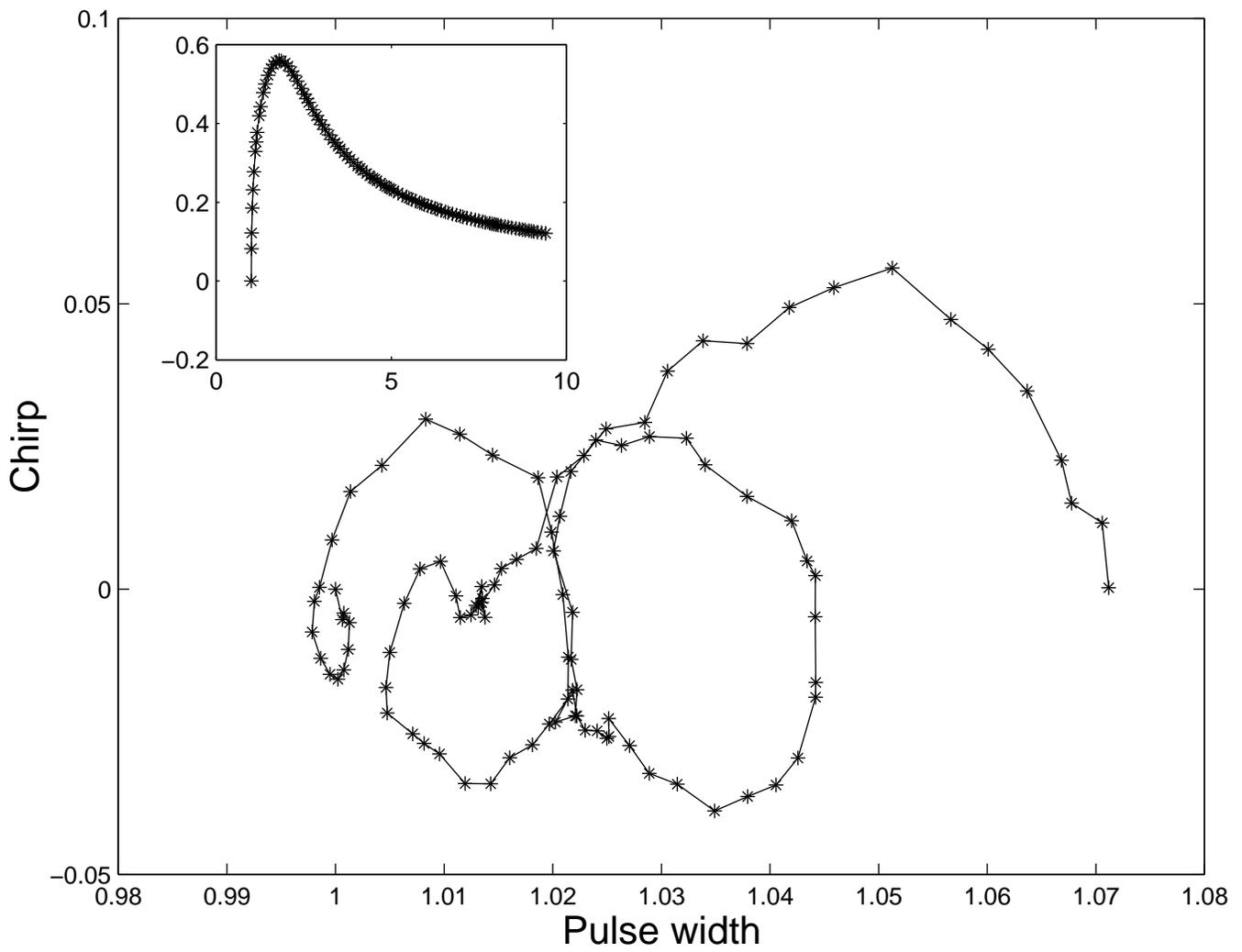
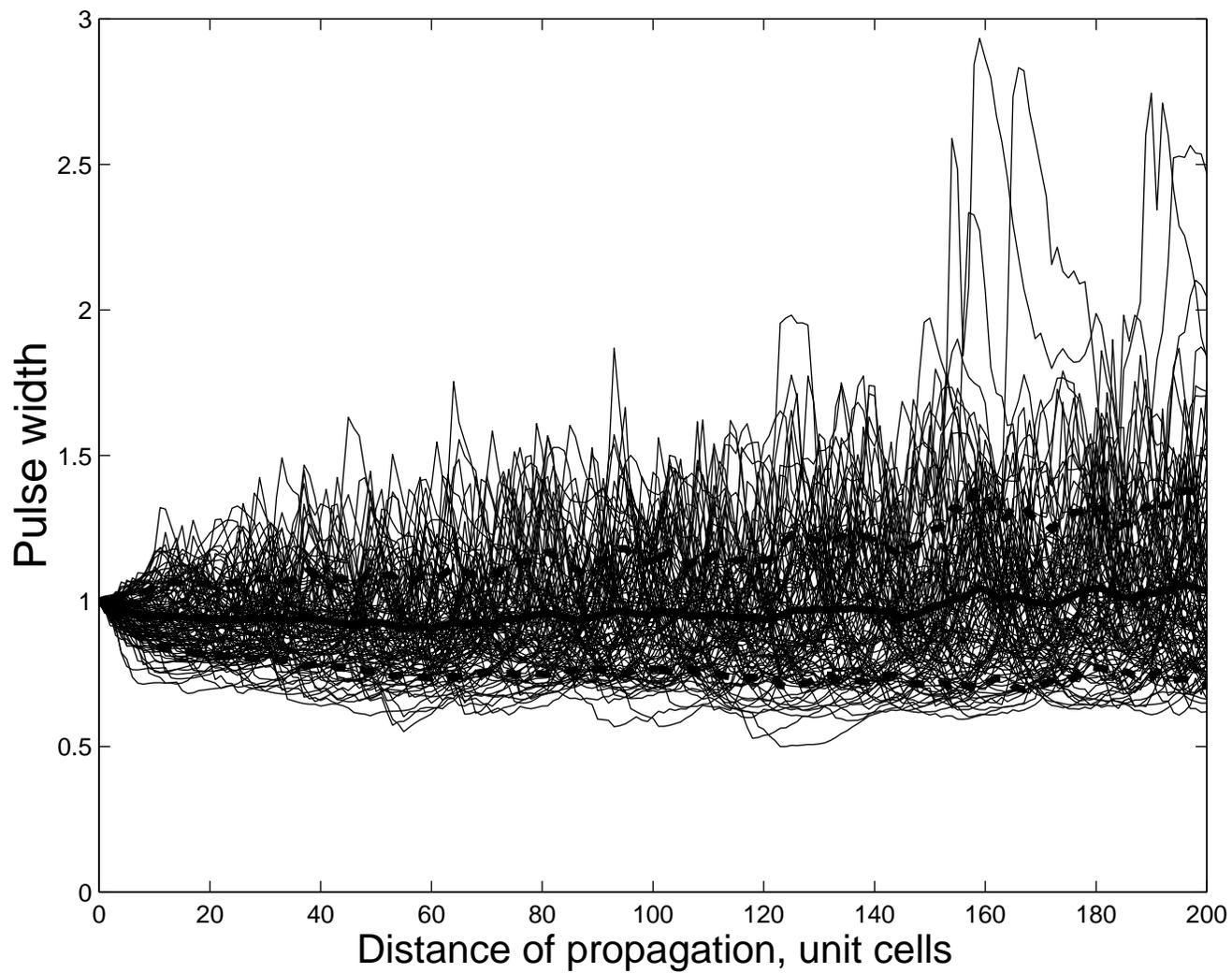
\end{document}